# Cherenkov Telescope Array: The Next Generation Gamma-ray Observatory

**The CTA Consortium[1], represented by Rene A. Ong[2]**

The Cherenkov Telescope Array (CTA) will be the next-generation gamma-ray observatory, investigating gamma-ray and cosmic ray astrophysics at energies from 20 GeV to more than 300 TeV. The observatory, consisting of large arrays of imaging atmospheric Cherenkov telescopes in both the southern and northern hemispheres, will provide full-sky coverage and will achieve a sensitivity improved by up to an order of magnitude compared to existing instruments such as H.E.S.S., MAGIC and VERITAS. CTA is expected to discover hundreds of new TeV gamma-ray sources, allowing it to significantly advance our understanding of the origin of cosmic rays, to probe much larger distances in the universe, and to search for WIMP dark matter with unprecedented sensitivity in TeV mass range.

The development of CTA is being carried out by a worldwide consortium of scientists from 32 countries. Consortium scientists have developed the core scientific programme of CTA and institutes of the Consortium are expected to provide the bulk of the CTA components. The construction of CTA is overseen by the CTA Observatory that will in the future manage observatory operations, the guest observer programme, and data dissemination.

This talk will review the scientific motivation for CTA, focusing on the key science projects that form the core programme of research. The talk will outline the design of CTA, including the science drivers, overall concept, performance optimization, and array layouts. The current status of CTA, including sites, prototype telescope progress, and steps forward will also be described.

---

[1]See http://www.cta-observatory.org/consortium_authors/authors_2017_07.html for full author list

[2]Speaker, Department of Physics and Astronomy, University of California, Los Angeles, Los Angeles, CA 90095, USA, *E-mail*: rene@astro.ucla.edu

## 1. Introduction

The field of very high energy (VHE, E > 20 GeV) gamma-ray astronomy came of age in 1989 with the clear discovery of the Crab nebula as a steady emitter of TeV gamma rays [1]. During the 1990's, the VHE catalog grew slowly as a few new sources were clearly detected, but a number of other objects could not be reliably confirmed [2]. However, it became clear during this period that the field had great scientific potential and that arrays of imaging atmospheric Cherenkov telescopes offered a logical path forward towards exploring the VHE universe with high sensitivity [3]. This reasoning led to the development of new atmospheric Cherenkov telescope arrays (e.g. H.E.S.S., MAGIC and VERITAS) that have proven to be remarkably successful by discovering many new sources of VHE gamma rays and by making major progress in understanding the source mechanisms through better determination of their properties (e.g. spectrum, morphology and variability). Indeed, the successes of the imaging atmospheric Cherenkov telescope arrays, combined with excellent results from the Fermi and AGILE satellite instruments and from air shower experiments (e.g. ARGO-YBJ, Milagro, and Tibet AS-$\gamma$), have demonstrated that VHE particle acceleration is ubiquitous in the universe [4]. In addition, it is now appreciated that the field can both significantly advance our understanding of high-energy astrophysical phenomena but that it can also address important questions in particle physics and cosmology.

The Cherenkov Telescope Array (CTA) is designed to be the next major observatory operating in the VHE gamma-ray band [5]. It will build on the well-proven imaging atmospheric Cherenkov technique but will go much further in terms of performance than current instruments. The scientific potential for CTA is great – among the important questions to be addressed are the origin of cosmic rays, the strength of intergalactic radiation fields, and the nature of dark matter. CTA will also be the first open observatory in this waveband and the synergies between CTA and other multi-wavelength and multi-messenger facilities are expected to further enhance CTA's scientific scope and relevance to other scientific communities.

CTA was conceived and is being designed by the CTA Consortium (CTAC), a collaboration of more than 1400 scientists and engineers from 32 countries around the world. The Consortium has developed the primary science themes of CTA and Consortium institutes are expected to provide the bulk of the CTA components, including telescopes, cameras and software. The CTA Observatory (CTAO) was established in 2014 to provide the legal entity to oversee the CTA Project Office that manages the construction of CTA. Governed by a Council of country representatives, CTAO will be responsible for observatory operations and data management. During the last several years, the progress towards realization of CTA has been accelerating. The baseline design and core technologies are now established, several prototype telescopes have been completed and are undergoing testing, the two CTA sites have been selected, and a large portion of the required funding has now been identified. Thus, the project is well positioned for a construction start in 2018 and the turn-on of full operations by the middle of the next decade.

## 2. CTA Science

### 2.1 Scientific Themes

Very high energy gamma-ray astronomy is a relatively young field with great scientific potential. The current generation atmospheric Cherenkov telescopes (H.E.S.S., MAGIC, and VERITAS), along with the newly operational HAWC air shower experiment, have firmly established the field, discovering VHE radiation from more than 150 sources, comprising many source classes. A number of individual sources, both within and outside of our Galaxy, have been well-studied but there are many others that are not well-characterized or understood. It seems clear that our current knowledge represents just the tip of the iceberg in terms of the number of sources and source classes and in terms of our ability to confront the existing theoretical models. CTA will transform our understanding of the high-energy universe by discovering many hundreds of new sources, by measuring their properties with unprecedented accuracy, and also by exploring questions in physics of fundamental importance. The major scientific questions that can be addressed by CTA are the following, grouped into three broad themes:

Theme 1: Understanding the Origin and Role of Relativistic Cosmic Particles
- What are the sites of high-energy particle acceleration in the universe?
- What are the mechanisms for cosmic particle acceleration?
- What role do accelerated particles play in feedback on star formation and galaxy evolution?

Theme 2: Probing Extreme Environments
- What physical processes are at work close to neutron stars and black holes?
- What are the characteristics of relativistic jets, winds and explosions?
- How intense are radiation fields and magnetic fields in cosmic voids, and how do these evolve over cosmic time?

Theme 3: Exploring Frontiers in Physics
- What is the nature of dark matter? How is it distributed?
- Are there quantum gravitational effects on photon propagation?
- Do axion-like particles exist?

### 2.2 Core Programme

Over the lifetime of CTA, most of the available observation time will be divided into the Guest Observer (GO) Programme, where time will be awarded based on scientific merit, and a Core Programme of a number of major legacy projects. Director's Discretionary Time and host country reserved time will comprise the remaining time. The CTA Consortium has developed the Core Programme that consists of proposed Key Science Projects (KSPs) that are characterized by having an excellent science case and clear advance beyond the state of the art, the production of legacy data sets of high value to the wider community, and clear added value for the project to be done as a KSP rather than part of the GO Programme (e.g. because of the scale of the project or the expertise required in carrying it out). This Core Programme has been initially described in a special issue of Astroparticle Physics [6] and more recently in the document "Science with the Cherenkov Telescope Array" [7]. The proposed CTA Key Science Projects include: (i) Dark Matter Programme, (ii) Galactic Center Survey, (iii) Galactic Plane Survey, (iv) Large Magellanic Cloud Survey, (v) Extragalactic Survey, (vi) Transients, (vii) Cosmic-ray PeVatrons, (viii) Star Forming Systems, (ix) Active Galactic Nuclei, and (x) Clusters of Galaxies. A few highlights from these projects are described here, focusing on the surveys and the search for dark matter:

- The **Galactic Centre Survey** consists primarily of a deep (525 h) exposure with pointings on a small grid centered on Sgr A*; this exposure covers the central source, the centre of the dark matter halo, the primary diffuse emission and multiple supernovna remnant (SNR) and pulsar wind nebula (PWN) sources. An extended survey (300 h) of a 10º x 10º region around the Galactic centre would cover the edge of the Galactic bulge, the base of the Fermi Bubbles, the radio spurs and the Kepler SNR.
- The **Galactic Plane Survey** is a survey of the entire Galactic plane, with deeper exposure in the inner Galaxy and Cygnus region [8]. The survey will be a factor of 5-20 more sensitive than previous surveys carried out at very high energies and is thus expected to sample a much larger fraction of the log $N$ – log $S$ distribution of Galactic sources, as shown in Figure 1. The discovery of many hundreds of sources in the Galactic Plane Survey will be an important pathfinder for later GO proposals.
- The **Large Magellanic Cloud (LMC) Survey** will cover this star-forming galaxy in its entirety, resolving regions down to 20 pc in size and with sensitivity down to a luminosity of ~$10^{34}$ erg/s. Long-term monitoring of SN 1987A will be carried out, provided the source is detected in the first phase of the survey.
- The **Extragalactic Survey** will be the first wide-field (one-quarter of the sky, see Figure 2) survey of the VHE sky at high sensitivity. Aimed to provide an unbiased sample of galaxies (particularly active Galactic nuclei, AGN), the survey will also be sensitive to unexpected phenomena at high Galactic latitudes.
- The **Dark Matter Programme** is centered on the indirect search for dark matter via the weakly interacting massive particle (WIMP) annihilation signal [9]. As shown in Figure 2, the deep exposure of the Galactic centre region will allow CTA to reach a sensitivity to a thermal relic WIMP over a wide mass region, thus nicely complementing searches done at the Large Hadron Collider and by direct-detection experiments. Additional dark matter targets include dwarf spheroidal galaxies, the LMC and the Perseus cluster.

## 2.3 Synergies

CTA will have important synergies with many of the new generation of astronomical and astroparticle observatories. These synergies, described in more detail elsewhere [7], will be exploited in a number of ways, including joint or follow-up observations carried out via memoranda of understanding between CTA and other facilities as well as the active involvement of many scientists in the CTA Guest Observer Programme and in the analysis of CTA data available on the public archive. In the high-energy and very high-energy gamma-ray regime, CTA will be well-complemented by satellite experiments (currently Fermi-LAT [10] and DAMPE [11]) and by ground-based facilities including HAWC [12] and LHAASO [13].

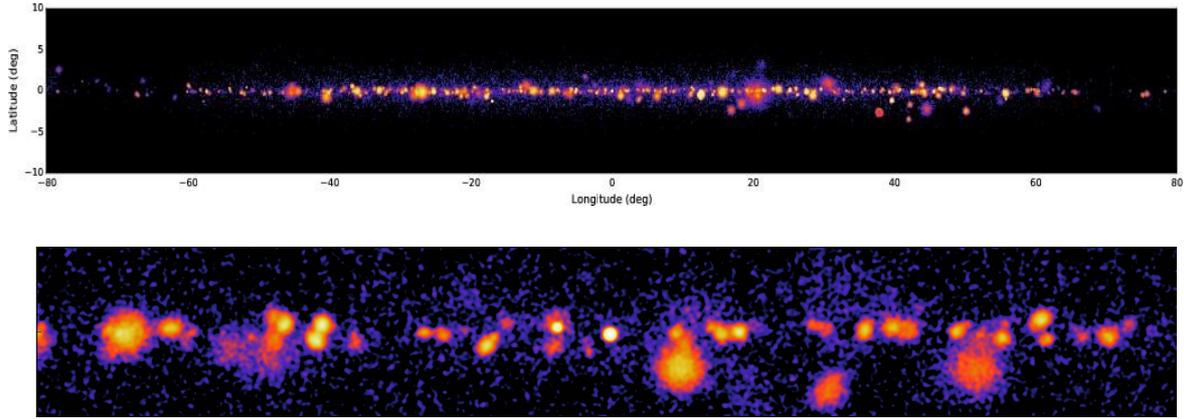

*Figure 1: Top: simulated CTA image of the Galactic plane for the inner region, adopting the proposed Galactic Plane Survey observation strategy and a source model that contains SNR and PWN populations and diffuse emission [7]. Bottom: a close-in view of a 20º region in Galactic longitude.*

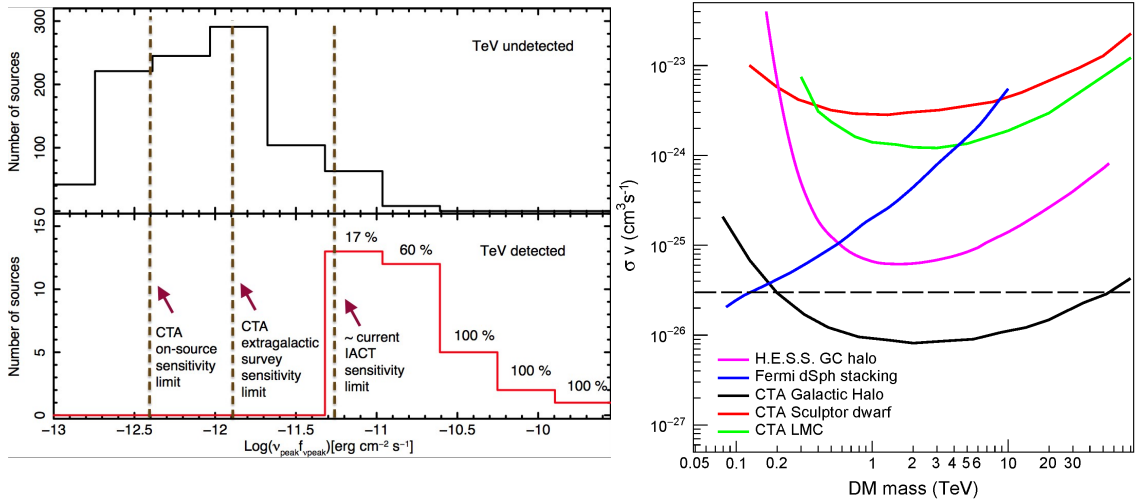

*Figure 2: Left: prediction for the number of blazars in the GeV-TeV domain – source counts versus peak synchrotron flux (see [7]). The upper panel shows predictions and the lower panel shows detected AGN. The vertical lines show the current sensitivity limit and the expected sensitivities for CTA. Right: CTA sensitivity to WIMP annihilation signature as a function of WIMP mass, for nominal parameters and for the multiple CTA observations; the dashed line shows the thermal relic cross section [7].*

## 3. CTA Design: Performance Goals, Concept, and Array Layouts

To achieve these broad science goals in a meaningful way, CTA must improve upon the performance of existing instruments in many areas simultaneously. The various performance goals, along with the science drivers that provide their impetus, are the following:

- **High sensitivity** (a factor of up to ten improvement over current experiments): impacts all science topics,
- **Wide Energy Coverage** (20 GeV to >300 TeV): low-energy sensitivity is needed to detect the most distant sources whose spectra are cut off from absorption on intergalactic radiation fields; very high-energy reach is needed to detect "PeVatron" sources that would help explain the origin of cosmic rays up to the knee in the spectrum,
- **Full-sky Coverage** (arrays in both hemisphers): allows the full characterization of the VHE universe and access to unique sources in both hemispheres,
- **Wide Field-of-View** (~8°): permits more rapid surveys and better study of extended sources,
- **Excellent Resolution** in angle (few arc-minutes) and energy (~10%): permits good reconstruction of source morphology and spectra, and
- **Rapid Response** (~20 s slewing): enables rapid follow up of transient sources.

To meet these performance goals, CTA will extend the atmospheric Cherenkov technique to its logical next level, by deploying large arrays of telescopes that cover an area on the ground that is significantly larger than the Cherenkov light pool. Compared to the existing instruments consisting of several telescopes separated by about 100m, the larger number of telescopes and the larger area covered by CTA will result in: i) a much higher rate of showers contained within the footprint of the array, ii) a better sampling of the showers from different viewing angles that will greatly improve the shower reconstruction and the cosmic-ray background rejection, and iii) a lower energy threshold since the central part of the shower (with the highest Cherenkov photon density) generally falls within the array.

To achieve the goal of wide energy range within cost constraints leads to the logical choice of a graded array of telescopes of different sizes. In CTA, the **lowest energies** are covered by four large-sized telescopes (LSTs) that are capable of detecting gamma rays down to 20 GeV. The **core energy range of 100 GeV to 10 TeV** is covered by an array of 25 (south) or 15 (north) medium-sized telescopes (MSTs), and the **highest energies** are covered by a several km$^2$ array of 70 small-sized telescopes (SSTs). To achieve fast-response to low-energy transients such as gamma-ray bursts, the LSTs will incorporate very rapid slewing. Conversely, to achieve a wide field-of-view for surveys and extended Galactic sources, the MSTs and SSTs will employ wide-field cameras. To realize full-sky coverage, CTA arrays will be deployed in both hemispheres. The small-sized telescopes are only planned for the southern array because the highest energies are most relevant for the study of Galactic sources.

The layout of the telescopes in the CTA arrays has been determined over a number of years by a multi-step process starting with semi-analytic estimates and continuing with large-scale simulations that include full shower and detector modeling. The latest (2017) simulations incorporate site-dependent effects (including altitude, geomagnetic field, and telescope positioning constraints) to assess the performance attributes of CTA. Figure 3 shows the current baseline array layouts for the southern and northern CTA sites resulting from this optimization process [14]. The simulations, coupled with the expected attributes of the various types of telescopes currently being designed and prototyped (see below), indicate that CTA will meet the ambitious performance goals described above [15].

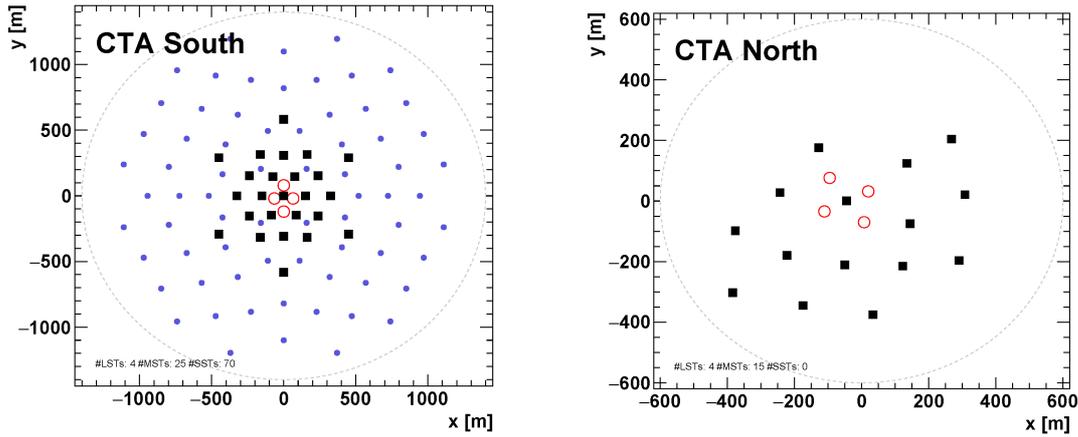

*Figure 3: Possible layouts for the baseline arrays for CTA South (left) and CTA North (right) [14]. The LSTs are identified by the red cirles, the MSTs by the black squares, and the SSTs by the blue diamonds.*

## 4. Current Status of CTA

### 4.1 CTA Sites

CTAO activities will be carried out at the two CTA array sites and at the CTA Headquarters (HQ) and Science Data Management Centre (SDMC). Pending successful completion of hosting agreements, the CTA HQ will be hosted at the INAF site in Bologna, Italy and the CTA SDMC will be on the DESY campus in Zeuthen, Germany. Following a lengthy process that included detailed assessment and external review, the CTA Resource Board (a precursor to the CTA Council) selected the following two sites to host CTA arrays:

- South: European Southern Observatory (ESO) Paranal site in Chile
- North: Instituto de Astrofisica de Canarias (IAC) Roque de los Muchachos Observatory site in La Palma, Spain

Activities to prepare the sites are well underway in both hemispheres. Technical and infrastructure studies are being carried out in the context of the Royal Institute of British Architects (RIBA) process [16]. CTA is currently in the developed design phase (RIBA-3) and is approaching the technical design phase (RIBA-4). Specific activities include power, lightning, geotechnical, ground investigation, and general infrastructure (roads, buildings, foundations, etc.) studies. On La Palma, the construction of the first prototype LST has started with the completion of its foundation. This prototype is expected to become the first LST in the northern CTA array.

### 4.2 Prototype Telescopes

Extensive work has been carried out within the CTA Consortium over a number of years to prototype the hardware and software for all three telescope types. This work builds on the successes and experiences of the current generation of imaging atmospheric Cherenkov telescopes, but it also makes use of new techniques. For example, in the telescope design, both single mirror (based on the traditional Davies-Cotton, or DC, design) and dual mirror (based on the Schwarzschild-Couder, or SC, design) approaches are being developed. For the photosensors in the cameras, both photomultiplier tubes (PMTs) and Silicon photomultipliers (Si-PMs) are being evaluated. In all camera designs, the read-out electronics (typically using 1 GS/s high-speed sampling ASICs) are contained in the focal-plane box.

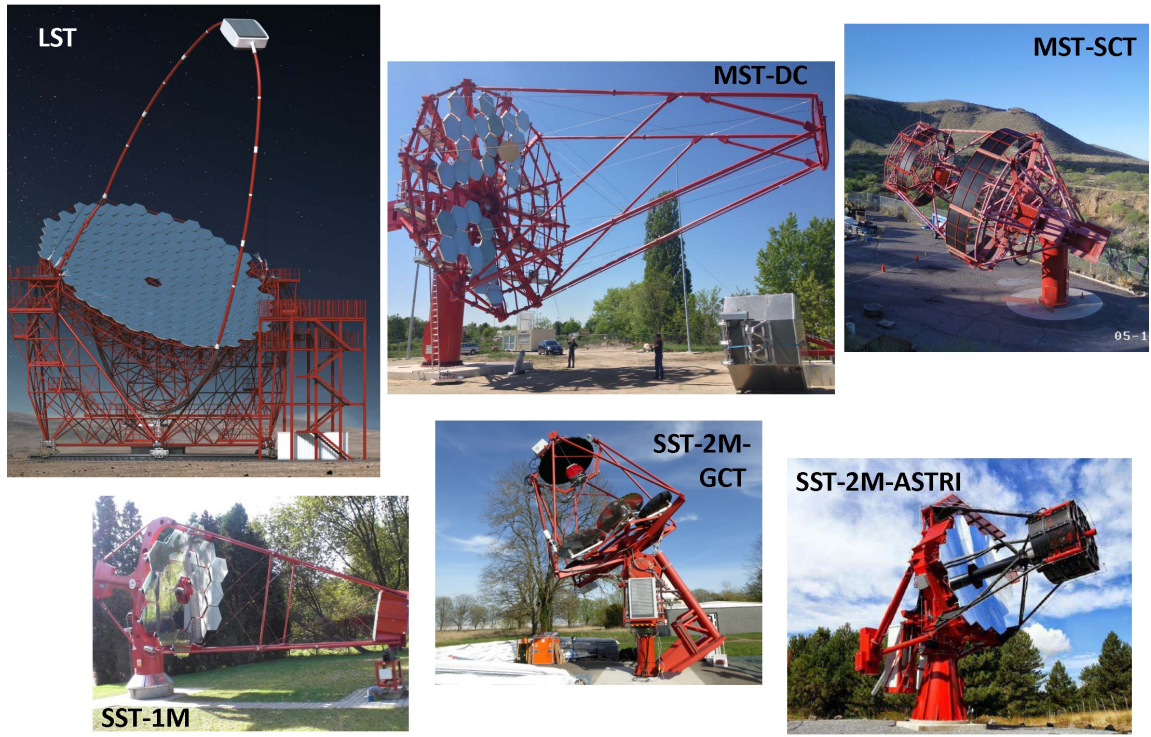

*Figure 4: Prototype telescopes being developed for CTA. Top row (left to right): LST design, MST-DC in Germany, MST-SCT in USA. Bottom row (left to right): SST-1M in Poland, SST-2M-GCT in France, and SST-2M-ASTRI in Italy. See text for details.*

Figure 4 shows recent photos of the various prototype CTA telescopes (and a conceptual drawing for LST). For the LST, the requirement of large mirror area to reach the lowest gamma-ray energies has led to a single mirror design using a 23 m diameter parabolic reflector. This very large telescope will use PMTs and will have rapid re-positioning capabilities to catch transient sources such as gamma-ray bursts. For the MST, two designs are being considered. A single mirror DC design has been developed at a site in Adlershof, Germany that makes use of a 12 m diameter dish with a focal length of 16 m and a PMT camera. Two read-out schemes are being prototyped that make use of 250 MS/s Flash-ADCs with digital storage and 1 GS/s ASICs. A dual mirror SC MST prototype is being built at the Whipple Observatory in Arizona, USA that will employ a 9.7 m primary mirror and a compact high-resolution camera using Si-PMs [17]. For the SST, three approaches are being considered, with each having a primary mirror size of 4 m diameter and cameras using Si-PMs. Two of these use the SC design: the SST-2M-ASTRI prototyped at Serra La Nave, Sicily, Italy [18] and SST-2M-GCT in Meudon, France [19]. The third SST prototype, SST-1M, is being developed in Krakow, Poland and makes use of the DC design [20]. As of the writing of these proceedings (June 2017), completed, or nearly completed, prototype structures exist for all telescope designs, except the LST, and prototype cameras are close to completion for all telescope types.

### 4.3 Steps Forward

Although a very large amount of preparatory work has already been carried out, a great deal remains to be done to make CTA a reality. With the general scientific themes for CTA firmly established, the CTA Consortium is now focused on: 1) readying the telescope hardware for pre-production and eventual construction, 2) assisting the observatory in the prepration of software pipelines and building up the analysis expertise through data challenges, and 3) developing the linkages with other communities to best capitalize on the observations carried out in the Core Programme. The CTA Project Office, governed by the CTA Observatory, is ramping up in size and is: 1) preparing the sites for initial deployment, 2) putting into place the procedures for pre-production and production readiness reviews and for in kind contribution agreements, and 3) developing the integrated construction and verification management plan.

The estimated cost for the *baseline implementation* of CTA, consisting of 99 telescopes (4 LSTs + 25 MSTs + 70 SSTs) in the southern array and 19 telescopes (4 LSTs + 15 MSTs) in the northern array is 400 MEuro (including FTE's). A very positive feature of the CTA design is that the arrays are scalable and excellent science can be carried out before the baseline configuration is reached. Considering various inputs, including the expected flow of funding from the various partner countries and a scientific/technical assessment made by the Consortium, the CTA Council defined in 2016 a *threshold implementation* that will require 250 MEuro (including FTE's) and that will allow CTA to initiate the formal start of construction. It is anticipated that the committed funding will exceed this threshold by late 2017 and that CTA construction will start in 2018. Additional funding that is committed in the future will be used to allow CTA to reach the baseline implementation.

**Acknowledgments**
We gratefully acknowledge financial support from the agencies and organizations listed here: http://www.cta-observatory.org/consortium_acknowledgments.

**References**

[1] T.C. Weekes *et al.*, *Observation of TeV Gamma Rays from the Crab Nebula using the Atmospheric Cherenkov Imaging Technique*, Astrophys. J. **342**, 379 (1989).

[2] See, for example, the proceedings of *Towards a Major Atmospheric Cherenkov Detector V* (Kruger Park), ed. O.C. de Jager (1997).

[3] See, for example, Rene A. Ong, *Very High-Energy Gamma-Ray Astronomy*, Physics Reports **305**, 93 (1998).

[4] For a good compilation of review articles covering VHE gamma-ray astronomy, see, for example, http://tevcat.uchicago.edu/reviews.html.

[5] B.S. Acharya *et al.*, *Introducing the CTA Concept*, Astroparticle Phys. **43,** 3 (2013).

[6] *Seeing the High-Energy Universe with the Cherenkov Telescope Array*, ed. J. Hinton, S. Sarkar, D. Torres, and J. Knapp, Astroparticle Phys. **42**, 1 (2013).

[7] The CTA Consortium, *Science with the Cherenkov Telescope Array*, to be submitted, August 2017.

[8] The CTA Consortium, represented by R. Zanin, *Observing the Galactic Plane with the Cherenkov Telescope Array*, in these proceedings (2017), PoS(ICRC2017)740 (2017).

[9] The CTA Consortium, represented by A. Morselli, T*he Dark Matter Programme of the Cherenkov Telescope Array*, in these proceedings, PoS(ICRC2017)921 (2017).

[10] W.B. Atwood *et al.*, The Large Area Telescope on the Fermi Gamma-ray Space Telescope Mission, Astrophys. J. **697**, 1071 (2009).

[11] J. Chang, *Dark Matter Particle Explorer: The First Chinese Cosmic Ray and Hard $\gamma$-ray Detector in Space*, Chinese Space Science **34**, 550 (2014).

[12] A.J. Smith *et al.*, *HAWC: Design, Operation, Reconstruction, and Analysis*, Proc. of the 34[th] Int. Cosmic Ray Conf. (The Hague), arXiv: 1508.05826 (2015).

[13] G. Di Sciascio *et al.*, *The LHAASO Experiment: from Gamma-ray Astronomy to Cosmic Rays*, Proc of the CRIS 2015 Conference (Gallipoli, Italy), arXiv: 1602:07600 (2015).

[14] P. Cumani *et al*, for the CTA Consortium, *Baseline Telescope Layouts of the Cherenkov Telescope Array*, in these proceedings, PoS(ICRC2017)811 (2017).

[15] G. Maier *et al.*, for the CTA Consortium, *Performance of the Cherenkov Telescope Array*, in these proceedings, PoS(ICRC2017)846 (2017).

[16] See https://www.ribaplanofwork.com/.

[17] V. Vassiliev *et al.*, for the CTA SCT Project, *Prototype 9.7m Schwarzschild-Couder Telescope for the Cherenkov Telescope Array: Project Overview*, in these proceedings, PoS(ICRC2017)838 (2017).

[18] M.C. Maccarone *et al.*, for the CTA ASTRI Project, *ASTRI for the Cherenkov Telescope Array,* in these proceedings, PoS(ICRC2017)855 (2017).

[19] H. Sol *et al*, for the CTA GCT Project, *Observing the Sky at Extremely High Energies with the Cherenkov Telescope Array: Status of the GCT Project*, in these proceedings, PoS(ICRC2017)822 (2017).

[20] I. Al Samarai *et al*, for the CTA SST-1M Project, *Performance of a Small Size Telescope (SST-1M) Camera for Gamma-ray Astronomy with the Cherenkov Telescope Array*, in these proceedings, PoS(ICRC2017)961 (2017).